# Spatiotemporal accessible solitons in fractional dimensions

Wei-Ping Zhong[1,2 *], Milivoj R. Belić[2], Boris A. Malomed[3], Yiqi Zhang[4], and Tingwen Huang[2]

[1] Department of Electronic and Information Engineering, Shunde Polytechnic, Guangdong Province, Shunde 528300, China

[2] Texas A&M Univsersity at Qatar, P.O. Box 23874 Doha, Qatar

[3]Department of Physical Electronics, School of Electrical Engineering, Faculty of Engineering, Tel Aviv University, Tel Aviv 69978, Israel

[4] Key Laboratory for Physical Electronics and Devices of the Ministry of Education & Shaanxi Key Lab of Information Photonic Technique,

Xi'an Jiaotong University, Xi'an 710049, China

*Corresponding author: zhongwp6@126.com

**Abstract**：We report solutions for solitons of the "accessible" type in globally nonlocal nonlinear media of fractional dimension (FD), *viz*., for self-trapped modes in the space of effective dimension $2 < D \leq 3$ with harmonic-oscillator potential whose strength is proportional to the total power of the wave field. The solutions are categorized by a combination of radial, orbital and azimuthal quantum numbers $(n,l,m)$. They feature coaxial sets of vortical and necklace-shaped rings of different orders, and can be exactly written in terms of special functions that include Gegenbauer polynomials, associated Laguerre polynomials, and associated Legendre functions. The validity of these solutions is verified by direct simulation. The model can be realized in various physical settings emulated by FD spaces; in particular, it applies to excitons trapped in quantum wells.

## 1. Introduction

Calculus in fractional-dimensional (FD) spaces has found various applications in modeling different physically relevant settings, such as, e.g., absorption spectra in quantum wells [1], non-crystalline solids [2], impurity levels in semiconducting heterostructures [3], and excitons in quantum wells [4]. Recently, FD propagation equations of the Schrödinger type were utilized for modeling optical solitons, which in this context are sometimes referred to as the fractional optical solitons [5-9]. In particular, FD nonlinear Schrödinger equations can be emulated by means of their integer-dimensional counterparts with a singular spatial modulation of the local strength of the cubic self-focusing nonlinearity [10]. On the other hand, linear electrodynamics in FD spaces was formulated in Ref. [11]. Fractional calculus, which was originally developed in a more abstract form, offers various tools in dealing with Laplacians of fractional dimension, ranging from the quantum Riesz fractional derivative involving Levi flights [9] to the multidimensional Stillinger formulation [12] in non-integer-dimensional spaces [4,13,14]. The latter approach is directly relevant to the fractional dimension $2 < D \leq 3$ which is considered in this work.

Temporal and spatial self-localization of wave packets is a fundamental topic in linear [15] and nonlinear [16] optics alike. Spatiotemporal solitary waves, also called the light bullets (LBs) [16], are self-sustained modes which are localized both spatially and temporally [17,18], due to the simultaneous balance of diffraction and dispersion by the material nonlinearity. Three-dimensional (3D) spatiotemporal localization was experimentally demonstrated in arrays of transversely coupled waveguides, i.e., in a semi-discrete setting [19]. Recently, the formation of (2+1)D spatial solitons due to the saturation of the cubic self-focusing was reported too [20].

Spatial solitons which may exist in nonlocal nonlinear media have drawn much interest lately [21-23]. In particular, a possibility to realize the so-called "accessible solitons" (linear modes trapped in a harmonic-oscillator (HO) potential, whose strength is proportional to the total power of the light beam) [24]) was demonstrated in the limit of strongly nonlocal nonlinearity [23]. A long-range nonlocal nonlinear response of the "accessible" type was demonstrated in nematic liquid crystals [25,26].

In this work, we consider solitons produced by the FD Schrödinger equation with the "accessible" nonlocal nonlinearity, represented by the self-induced HO potential, which makes it possible to write the underlying Schrödinger equation in spherical coordinates (following the convention widely adopted in physics literature, we apply word "soliton" to robust self-trapped modes, even if they do not comply with the rigorous mathematical definition of solitons in integrable systems).

Due to the universality of the HO potential, this model is relevant for various physical applications [1-4, 11-14]. The FD accessible solitons are strongly localized, similar to Gaussian wave functions in quantum mechanics with an external HO potential. Several types of such modes are constructed here, including coaxial sets of vorticity rings and necklace rings.

It is relevant to mention that an analysis of FD accessible solitons in two spatial coordinates was recently published in Ref. [9] by some of the present authors. The present paper extends the analysis to three spatial coordinates in the FD space. Thus, the coordinates adopted in this paper are spherical, whereas in Ref. [9] they were polar. This difference leads to an essential change in the FD Laplacian [14] and, consequently, in the results (even if the solutions in both cases can be expressed in terms of similar special functions). Accordingly, the solutions here are characterized by a set of three quantum numbers, instead of two in Ref. [9].

The FD model considered in the present work is formulated in Section 2. Exact solutions of the model are presented in Section 3. Detailed analysis of the solutions and their comparison with numerical results is given in Section 4. The paper is concluded by Section 5.

## 2. The model

The model is based on the Schrödinger equation for the complex field amplitude $u(\zeta, r, \theta, \varphi)$ [13]:

$$i\frac{\partial u}{\partial \zeta} + \frac{1}{2}\nabla_D^2 u - sr^2 u = 0, \qquad (1a)$$

where $\zeta$ is the evolution variable, and $(r, \theta, \varphi)$ is the set of spherical coordinates varying in the usual intervals $0 \leq r < \infty$, $0 \leq \theta < \pi$, and $0 \leq \varphi < 2\pi$, while $\nabla_D^2$ is the FD Laplacian, relevant for the FD space with dimensions $2 < D \leq 3$, which is considered in the present work. Thus, the model is essentially based on the quantum harmonic oscillator in a space of non-integer dimension, $D$, which is applicable to many physical settings, see Refs. [4,10] and references therein. In this work, we use the FD Euler-Lagrange equations written in spherical coordinates, which is consistent with the form of the HO potential, thus being relevant to the above-mentioned physical settings [1-4,11-14]. The exact form of the FD Laplacian is [14]:

$$\nabla_D^2 = \frac{1}{r^{D-1}}\frac{\partial}{\partial r}\left(r^{D-1}\frac{\partial}{\partial r}\right) + \frac{1}{r^2 \sin^{D-2}\theta}\frac{\partial}{\partial \theta}\left(\sin^{D-2}\theta\frac{\partial}{\partial \theta}\right) + \frac{1}{r^2 \sin^2\theta}\left(\frac{\partial^2}{\partial \varphi^2} + \frac{D-3}{\tan\varphi}\frac{\partial}{\partial \varphi}\right). \qquad (1b)$$

Further, parameter $s$ in Eq. (1a) is proportional to the total norm of the wave field, which introduces the HO potential in the model for accessible solitons [24]:

$$s = s_0 \int_0^\infty r^{D-1} dr \int_0^\pi \sin^{D-2}\theta d\theta \int_0^{2\pi} \sin^{D-3}\varphi d\varphi |u(r,\theta,\varphi)|^2, \qquad (1c)$$

with the scaling coefficient $s_0$. The expression for the integral norm in Eq. (1c) is consistent with the form of Laplacian (1b) [14]. Without Eq. (1c), which introduces the global nonlinearity, Eq. (1a) with given constant $s$ provides for the model of the quantum-mechanical HO in the FD space. Obviously, Eq. (1) reduces to the usual 3D model when $D = 3$.

## 3. Spherical eigenmodes in the fractional-dimensional space

Similar to the 3D Schrödinger equation, Eq. (1) admits a solution with separated variables, $u(\zeta, r, \varphi, \theta) = R(\zeta, r)Y(\theta, \varphi)$. Plugging this ansatz into Eq. (1), one obtains

$$\frac{\partial^2 Y}{\partial \theta^2} + \frac{(D-2)\cos\theta}{\sin\theta}\frac{\partial Y}{\partial \theta} + L(L+1)Y + \frac{1}{\sin^2\theta}\left(\frac{\partial^2 Y}{\partial \varphi^2} + \frac{D-3}{\tan\varphi}\frac{\partial Y}{\partial \varphi}\right) = 0, \quad (2a)$$

$$i\frac{\partial R}{\partial \zeta} + \frac{D-1}{2r}\frac{\partial R}{\partial r} + \frac{1}{2}\frac{\partial^2 R}{\partial r^2} - sr^2 R - \frac{L(L+1)}{2r^2}R = 0, \quad (2b)$$

where the separation constant is $L(L+1) = (1/4)[(2l-1)(2l+3) + D(4-D)]$, with $l = 0,1,2,\cdots$, which, similar to the standard quantum mechanics, may be defined as the orbital quantum number. Angular equation (2a) admits further separation of variables, $Y(\theta,\varphi) = \Phi(\varphi)\Theta(\theta)$, admitting solutions in terms of Gegenbauer polynomials, $C_m^{|D-3|/2}(\cos\varphi)$, for $\Phi(\varphi)$ [27-29] and associated Legendre functions, $P_l^M(\cos\theta)$, for $\Theta(\theta)$:

$$\Phi(\varphi) = C_m^{|D-3|/2}(\cos\varphi), \quad \text{for } D \neq 3, \quad (3a)$$

$$\Theta(\theta) = (\cos^2\theta - 1)^{\frac{3-D}{4}} P_l^M(\cos\theta), \quad (3b)$$

where $m = 0,1,2,\cdots$ is the "magnetic" quantum number, alias the topological charge and $M \equiv m + (1/2)(D-3)$ These solutions are existing under the constraint $l \geq M$, which goes over to the usual constraint, $l \geq M$, in the case $D = 3$. We stress the fact that the solution given by Eq. (3a) is valid for $D \neq 3$.

Extending results available for the integer-dimensional space [23,24], one obtains the radial part of the FD accessible-soliton solution from Eq. (2b):

$$R(\zeta,r) = \frac{1}{\sqrt{w_0^3}}\left(\frac{r^2}{w_0^2}\right)^{\frac{2l+3-D}{4}} L_n^{(l+\frac{1}{2})}\left(\frac{r^2}{w_0^2}\right) e^{-\frac{r^2}{2w_0^2} - \frac{\zeta}{2}i(4n+2l-1)}, \quad (4a)$$

where $n = 0,1,2,\cdots$ is the radial quantum number, $L_n^{(l+1/2)}$ is the associated Laguerre polynomial [27], and the transverse width of the solutions is

$$w_0 \equiv 1/2\sqrt{s}. \quad (4b)$$

Using Eq. (4a) and the solution obtained for $Y(\theta,\varphi)$ when $D \neq 3$, the exact accessible-soliton solution of Eq. (1) can be written as

$$u(\zeta,r,\theta,\varphi) = \frac{k}{\sqrt{w_0^3}} C_m^{|D-3|/2}(\cos\varphi)(\cos^2\theta - 1)^{\frac{3-D}{4}} P_l^M(\cos\theta)\left(\frac{r^2}{w_0^2}\right)^{\frac{2l+3-D}{4}} L_n^{(l+\frac{1}{2})}\left(\frac{r^2}{w_0^2}\right) e^{-\frac{r^2}{2w_0^2} - \frac{\zeta}{2}i(4n+2l-1)}, \quad (5)$$

where the normalization constant is

$$k = \sqrt{\frac{2^{l+2-n}(2n+2l+1)!!}{\sqrt{\pi}n![(2l+1)!!]^2} \cdot \frac{|l+2m+D-3|!}{|2m+D-3||l-2m-D+3|!}}$$

and the factorial of the non-integer argument is realized as $\nu! \equiv \Gamma(\nu+1)$.

It is easy to see from Eq. (5) that $|u(z,r,\theta,\varphi)| \to 0$ at $|r| \to \infty$, i.e., the solution is indeed localized. Strictly speaking, so obtained solutions are unstable, because a deviation of the width from the above-mentioned value (4b) will break the stationary shape of the mode. Nevertheless, these solutions, even if perturbed, may persist over a sufficiently long propagation distance, which makes the analysis of these solutions relevant (see below). In the subsequent part of the paper, we display and discuss solutions given by Eq. (5) for different sets of quantum numbers $(n,l,m)$, with fixed initial width

$w_0 = 1$ and fixed FD, $D = 2.1$.

## 4. Analysis of the exact solutions and comparison with numerical results

For $m = 0$ (when $D \neq 3$, $C_0^{|D-3|/2}(\cos\varphi) \equiv 1$), analytical solution (5) yields a family of profiles with a multilayered structure, as shown in Figs. 1 and 2, for different $n$ and $l$. When $n = l = m = 0$, the solution amounts to the fundamental mode, since it has a spherical form for the integer-dimensional case ($D = 3$). However, due to the introduction of the FD ($D \neq 3$), the spherical structure is broken. Figure 1(a) displays a typical intensity distribution of the FD modes given by Eq. (5) with $D = 2.1$. It is seen that the fundamental FD solution is composed of two layers, and the intensity is zero at $x = y = 0$. For convenience, the field-amplitude distributions are drawn in the Cartesian $(x,y,\tau)$ coordinates. Figure 1(b) shows that the angular intensity, $|\Theta(\theta)|^2$, vanishes at $\theta = \pi$, as follows from Eq. (3b) for $D \neq 3$.

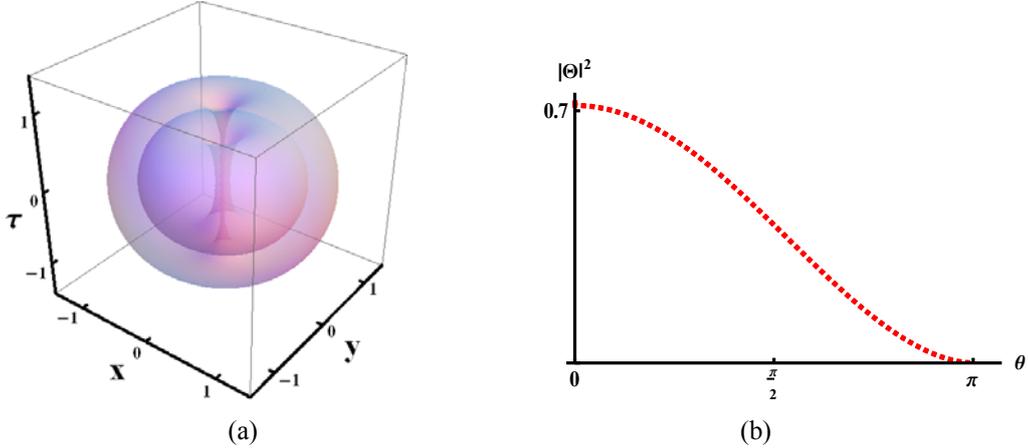

(a) (b)

Fig. 1. FD solutions for $D = 2.1$. (a) The structure of the fundamental solution with $n = l = m = 0$, shown (here and below) in terms of Cartesian coordinates $x, y, \tau$ corresponding to the spherical coordinates. (b) Intensity $|\Theta(\theta)|^2$ produced by Eq. (3b) for $l = 0$ and $m = 0$.

Next, we consider the solutions with $l \neq 0$ and $m = 0$ for $D = 2.1$. The solitons of this type are built of large inner rings and small capping ring-shaped objects. Generally, the larger $n$, the flatter the cap rings, while for the middle ring the trend is in the opposite. As $l$ increases, the number of rings decreases. There are $n+1$ embedded rings in the $(x,y)$ plane, and $2n+l+1$ layers along the $\tau$-axis. The intensity of the field attains maximum near the central ring.

The FD destroys the symmetry of the ring structure along the $\tau$-axis. It is worthy mentioning that structures featured by these axially-symmetric FD solutions resemble those reported for 3D gap solitons in the uniform self-defocusing medium, including a combination of an axially-periodic potential and a transverse HO trap [30].

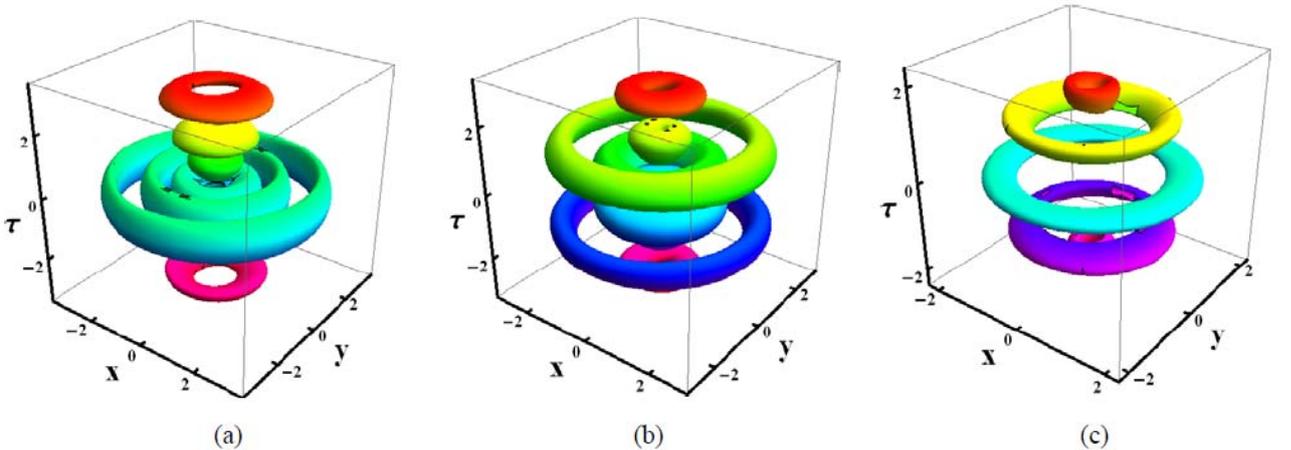

(a) (b) (c)

Fig. 2. Multilayered FD solutions for $m = 0$ and various sets of the two other quantum numbers: (a) $n = 2$, $l = 2$; (b) $n = 1$, $l = 3$; (c) $n = 0$, $l = 4$. The color is used here and in other figures only to distinguish different layers in the solutions.

Now, we consider solutions (5) with $m \neq 0$. In this case, they form ring-shaped chains of ellipsoids, which may be called asymmetric necklace ring solitons. Figure 3(a) displays a relatively simple example of the horizontal distribution for $l = m$ and different values of $n$, with different sizes of the necklaces. The smaller the magnetic quantum number $m$, the more complex shape featured by these solutions. In accordance with the necklace structure of solutions, the wave field vanishes along the $\tau$-axis, there being $2m$ "beads" in each necklace ring, and $n+1$ rings in the horizontal plane. Similar necklace-ring profiles may appear in the $(x, \tau)$ plane. A typical example for $m=1$, different $n$ and $l$ ($l \neq m$) is displayed in Fig. 3(b). In this case, the solution is composed of $2(n+1)l$ beads forming $n+1$ vertical rings with $l$ layers in the vertical direction, which get stretched in the vertical direction with the decrease in $l$.

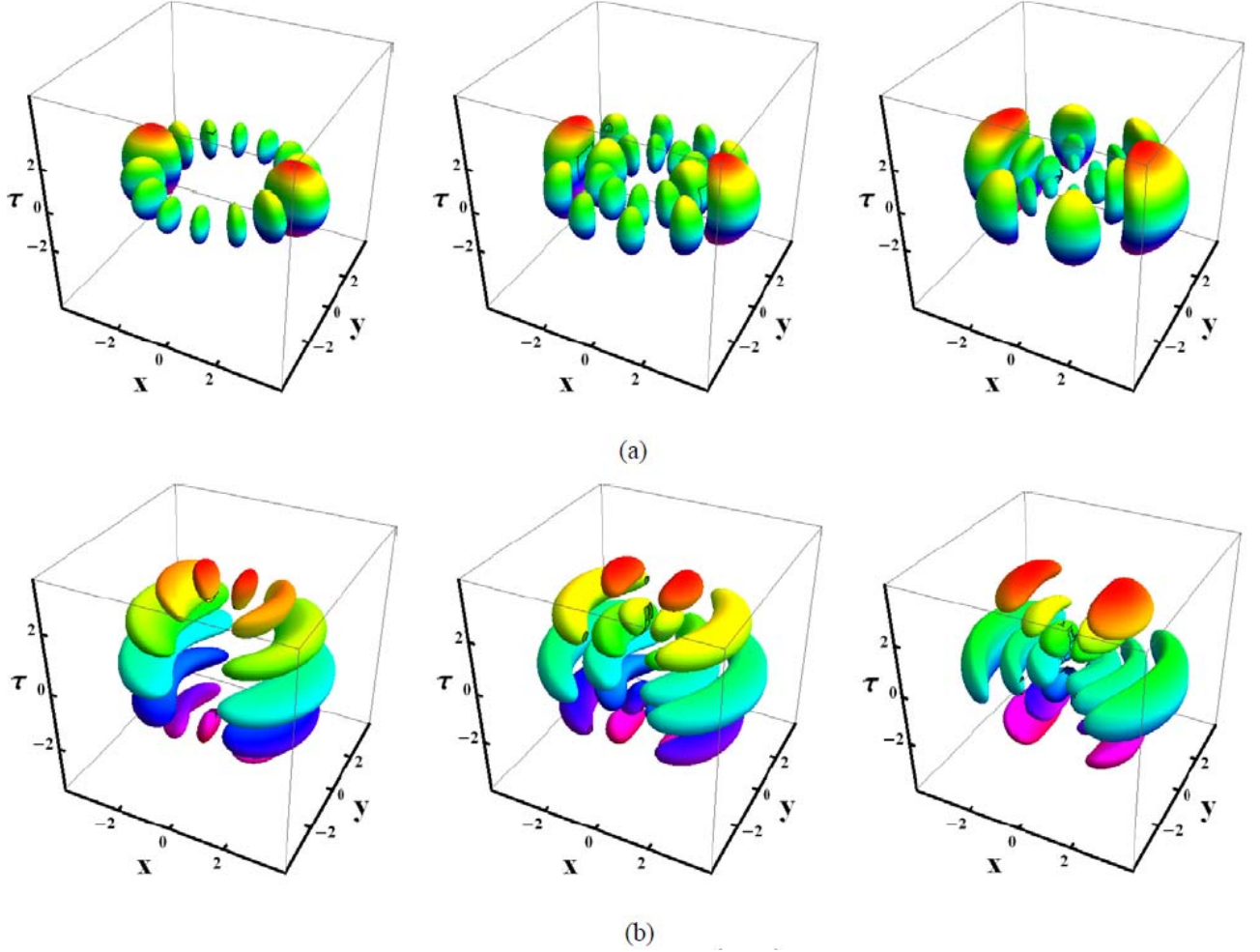

Fig. 3. Asymmetric necklace-ring solutions for various sets of quantum numbers $(n,l,m)$. (a) Horizontal ring with $l=m$ and different $n$: left $(n,l,m)=(0,7,7)$, center $(n,l,m)=(1,5,5)$, right $(n,l,m)=(2,3,3)$. (b) A vertical ring with $m=1$, $n=0$, and different $l$: left $(n,l,m)=(0,7,1)$, center $(n,l,m)=(1,5,1)$, right $(n,l,m)=(2,3,1)$. Recall that the color is used to distinguish different features in the intensity distribution, and is not related to the level of intensity.

In addition to the family of simple necklace ring modes shown in Fig. 3, solution (5) may also generate more sophisticated FD necklace-shaped solutions for $m \neq 0$. These patterns are composed of $2m(n+1)(l-m+1)$ beads, arranged into $n+1$ necklace rings in the $(x,y)$ plane, and $l-m+1$ tiers along the $\tau$-axis, as shown in Fig. 4, each ring containing $2m$ beads. In particular, Fig. 4(a), corresponding to $n=0$, features single three-, four-, and five-tier necklace rings. With increasing $m$, the beads get more stretched in the $\tau$ direction, and the entire necklace tends to merge into a uniform ring, cf. Fig. 2. Further, it is possible to demonstrate that in the multi-ring necklaces, the size of the ellipsoidal beads is the largest in the outermost ring, and the smallest in the innermost one, as seen in Fig. 4(b).

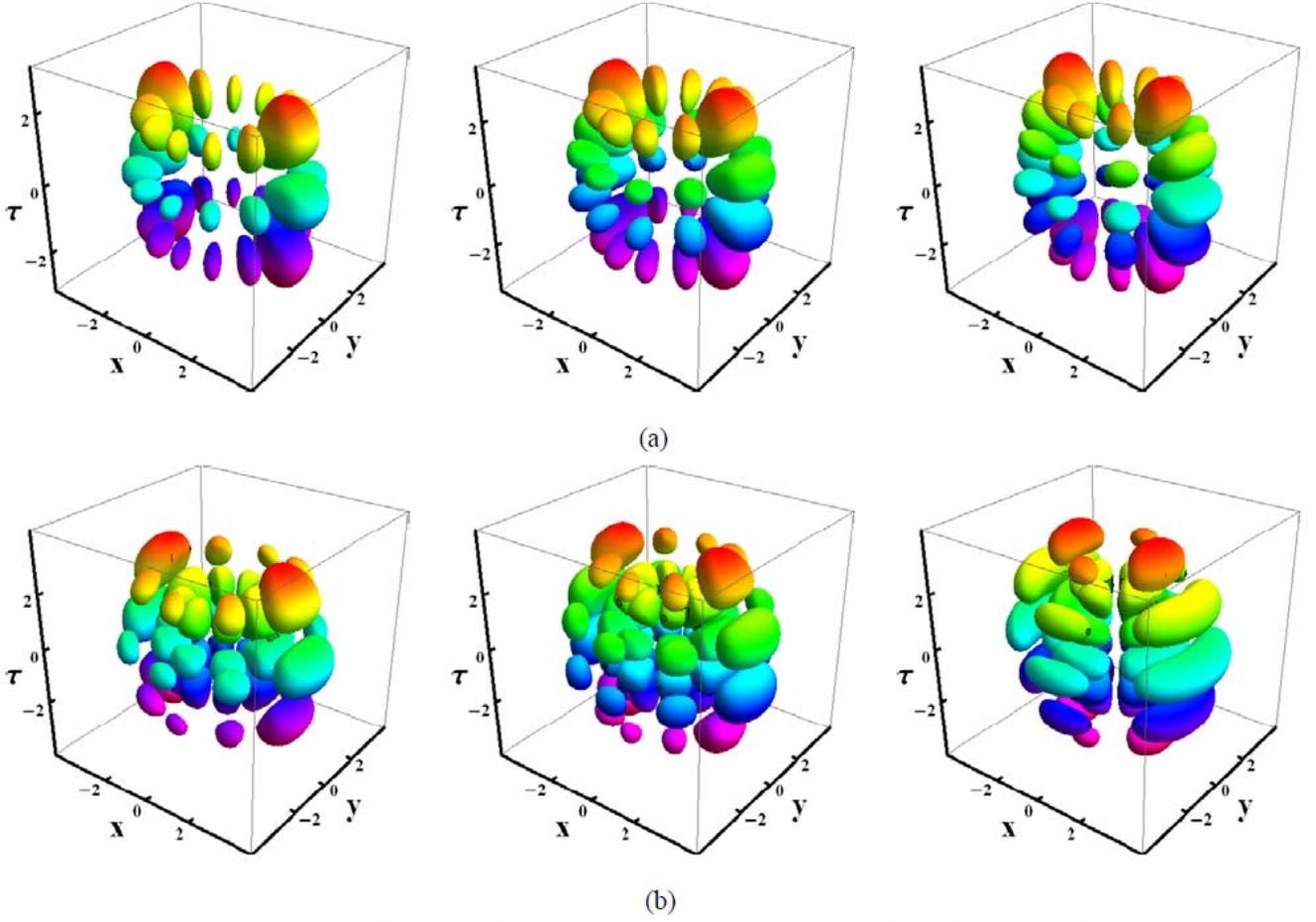

Fig. 4. Necklace solitons: (a) $n = 0$, $l = 8$ and $m = 6,5,4$ from left to right, respectively; (b) $n = 1$, $l = 6$ and $m = 4,3,2$ from left to right.

Finally, we address the validity of these solutions, which is verified by direct numerical simulations of Eq. (1) (performed with the help of the split-step beam propagation method, similar to that used in Ref. [31]). Simulations also confirm the dynamical stability of analytical solutions (5) by comparing them to their numerical counterparts, see an example in Figs. 5(a) and (b).

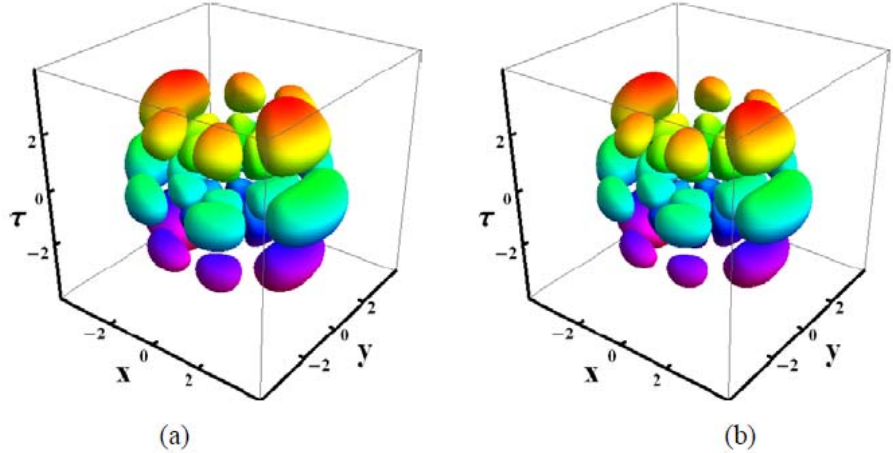

Fig. 5. Comparison of exact solution (5) with the numerically found counterpart. (a) The analytical solution given by Eq. (5); (b) results of numerical simulations of Eq. (1) at $\zeta = 80$ initiated by this exact solution, for the same case as in Fig. 4, except that $(n,l,m) = (1,5,3)$.

On the other hand, it was mentioned above that the FD accessible solitons cannot be entirely stable in the present model, as they are subject to the self-compression or expansion if condition (4b) is broken. Detailed results for the structural stability and dynamical behavior of the solutions at $w \neq w_0$ will be reported separately.

## 5. Conclusion

We have introduced a model that allows one to construct exact solutions for various types of "accessible solitons" with fractional dimension in globally nonlocal nonlinear media, which may find diverse physical realizations. The solutions were obtained by means of the separation of variables, which gives rise to the appearance of three types of special functions, *viz*., the Gegenbauer and associated Laguerre polynomials and associated Legendre functions in the solutions. The FD solutions display diverse shapes which do not appear in the usual 3D case, including vorticity-ring sets and necklace-shaped rings. These solutions are stable in a limited sense, over extended evolution intervals. The structural stability against deviations of the width from the special value given by Eq. (4b) will be considered elsewhere. Our results suggest that a search for FD self-similar solutions in other physical models, such as the Gross-Pitaevskii equation for Bose-Einstein condensates, is feasible and may be relevant as well.

**ACKNOWLEDGMENTS**

This work was supported by the National Natural Science Foundation of China (No. 61275001) and by the Natural Science Foundation of Guangdong Province, China (No. 2014A030313799). Work at the Texas A&M University at Qatar was supported by the NPRP 6-021-1-005 project with the Qatar National Research Fund (a member of Qatar Foundation). MRB acknowledges support from the Al Sraiya Holding Group.